\documentclass[journal]{IEEEtran}
\usepackage {graphicx}
\usepackage{subfigure}
\DeclareGraphicsExtensions{.eps,.pdf,.jpeg,.png}
\usepackage[cmex10]{amsmath}
\usepackage{amssymb}
\usepackage{cite}
\usepackage{multirow}
\usepackage{array} 
\begin{document}

\title{On Non-Orthogonal Multiple Access in Downlink}

\author{Xuezhi Yang,~\IEEEmembership{Senior Member,~IEEE }

\thanks{Xuezhi Yang is the inventor of several 4G fundamental techniques including  soft frequency reuse, scalable OFDM and random beam forming. His new invention, multi-level soft frequency reuse,  can improve the overall spectrum efficiency of 4G network by 30\% with near zero cost. He is  seeking the chance of cooperation with the industry. He is located in  Beijing, China  (email: yangxuezhi@hotmail.com) }
\thanks{}}

\markboth{}{}

\maketitle


\begin{abstract}

Non orthogonal multiple access (NOMA) in the downlink is discussed in this letter.  When combined with soft frequency reuse, NOMA is detrimental to user fairness by increasing the data rate of a near user at the cost of data rate of a far user.

\end{abstract}

\begin{IEEEkeywords}
Non orthogonal multiple access, downlink, soft frequency reuse, 5G.
\end{IEEEkeywords}


\section{Introduction}

Multiple access techniques are used to allow a  number of mobile users to share the same spectrum.  FDMA, TDMA, CDMA and OFDMA are used in the 1th to 4th generation mobile communication systems, all in an orthogonal way.  With the coming of 5G, non orthogonal multiple access (NOMA) became popular, both in the downlink \cite{DownlinkNOMA} and uplink\cite{UplinkNOMA}. 

It was once widely believed that superposition coding is the way to achieve maximum capacity, dominating orthogonal measures as time division (TD) or frequency division (FD)\cite{DavidTse, Andrea}. However, it was pointed out in \cite{MAC}  that TD, FD and superposition coding have the same capacity region under the constraint of sum power in multiple access channel (MAC). So NOMA is not an option for 5G uplink. 

In the downlink,  the two user model can be expressed as

\begin{eqnarray}
y_1&=&h_1(x_1+x_2)+n_1,\\
y_2&=&h_2(x_1+x_2)+n_2,
\end{eqnarray}
where $x_1$ and $x_2$ are signals of user one and two,  $h_1$ and $h_2$ are the channel gain , $n_1$ and $n_2$ are white noise, $y_1$  and $y_2$ are  the received signal.  

\begin{figure}[!ht]
\begin{center}
\includegraphics[width=3in]{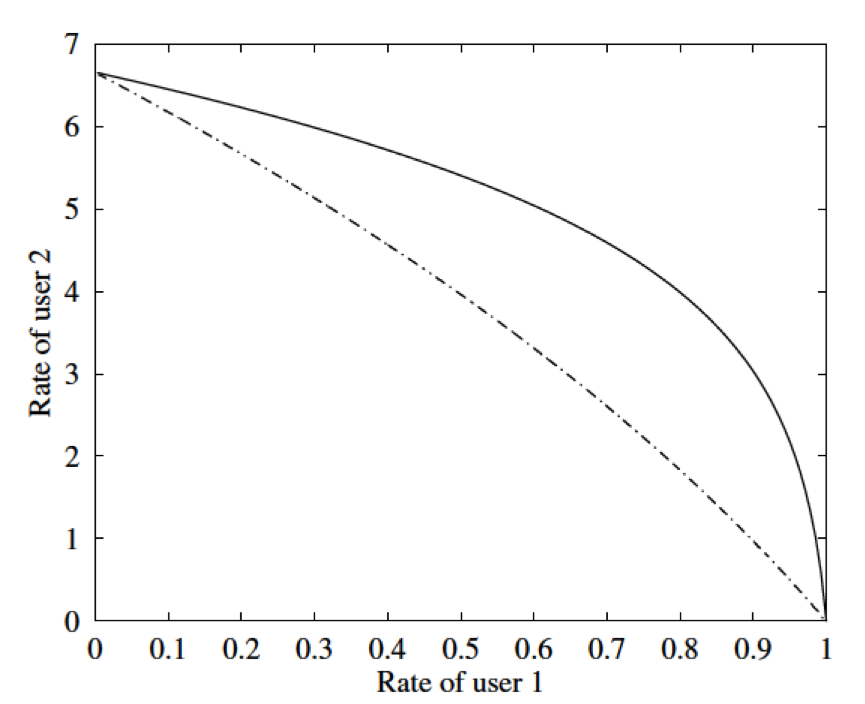}
\caption{The boundary of rate pairs (in bits/s/Hz) achievable by superposition coding (solid line) and orthogonal schemes (dashed line) for the two-user asymmetric downlink AWGN channel with the user SNRs equal to 0 and 20dB \cite{DavidTse}.}
\label{Fig-PTx2}
\end{center}
\end{figure}

In case of symmetric channel, i.e. $h_1=h_2$, such a broadcasting channel is equivalent to a MAC, then NOMA can achieve no capacity gain over orthogonal schemes. However, in an asymmetric channel,  i.e. $h_1\neq h_2$, NOMA will have a larger capacity region than OMA, as illustrated in Fig.  \ref{Fig-PTx2} \cite{DavidTse}, where the gain difference for the two users is 20dB. Especially, when the rate of user 1 is 0.9, only 10\% lower than the maximum rate, the rate of user 2 with NOMA is about 3 times  of that with OMA. This is regard as a great superiority  of NOMA over OMA. 

\section{Soft frequency reuse}

\begin{figure}[!ht]
\begin{center}
\includegraphics[trim=0.8in 0.5in 0.5in 0in, width=3in]{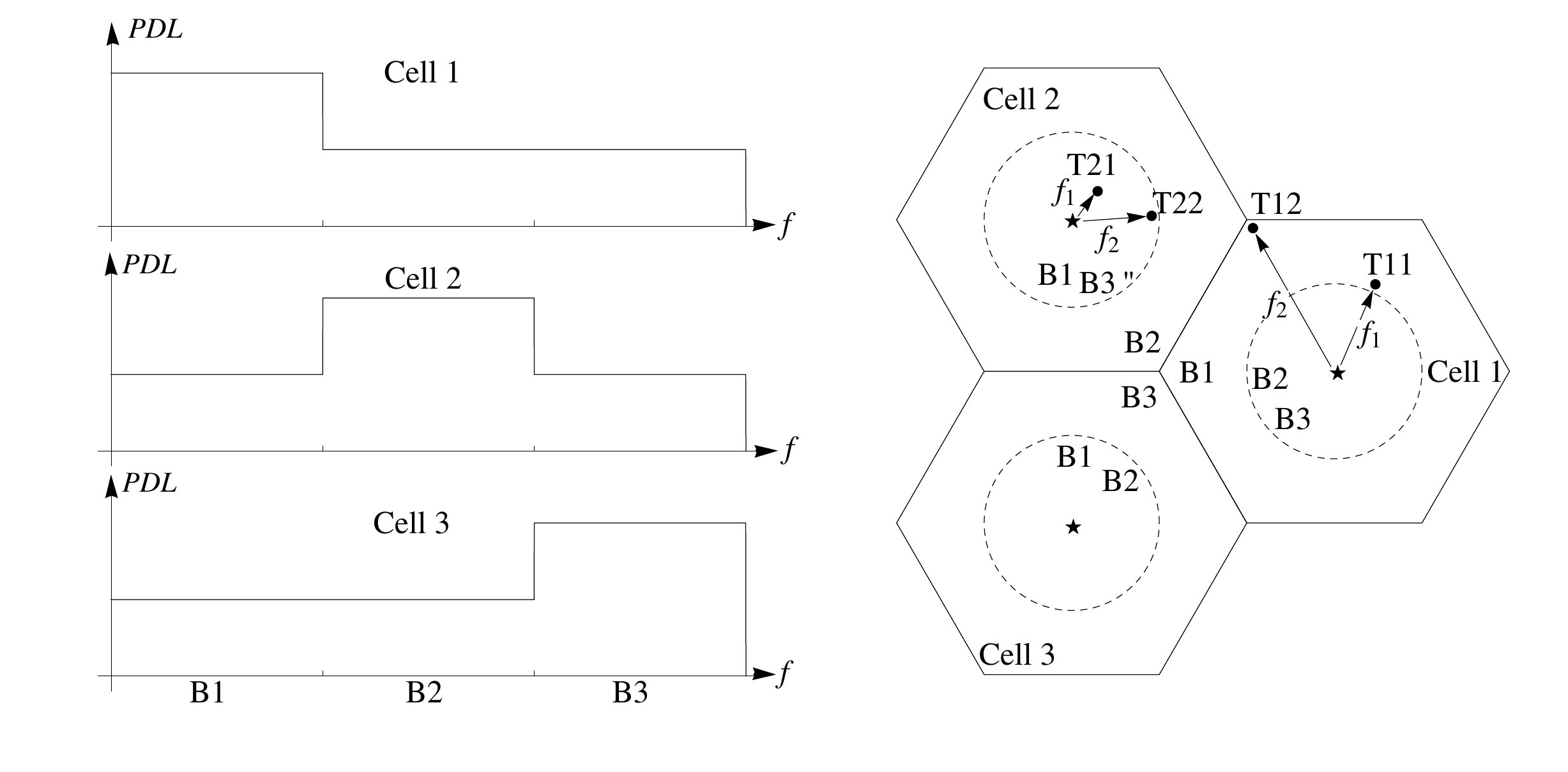}\\
(a)     \qquad   \qquad  \qquad   \qquad\qquad  (b)
\caption{Power density upper limit and coverage in  soft frequency reuse.}
\label{Fig-SFR}
\end{center}
\end{figure}

Orthogonal frequency division multiplexing (OFDM) was adopted in LTE to overcome the intra-cell  multi-user interference, a main drawback of code division multiple access (CDMA) in 3G.  Then, the inter-cell interference problem became the main obstacle to improve the user data rate, especially for cell edge users. Soft frequency reuse (SFR) \cite{YangSFRPatent2006, HuaweiSfr2005} was introduced to LTE to solve the inter-cell interference problem. In an example SFR configuration, as illustrated in Fig. \ref{Fig-SFR}, the whole bandwidth is divided into three parts, in which one part, called primary band, has a higher power density upper limit (PDL) than the other two, called secondary bands. The primary bands of adjacent cells are orthogonal to and do not interfere with each other. The rationality of SFR lies in that, even only a part of the whole bandwidth is used at cell edge, the data rate is increased compared to a reuse 1 scheme since the improvement in SINR is large enough to compensate the bandwidth loss \cite{HuaweiSfr2005}.  

\begin{figure}[!ht]
\begin{center}
\includegraphics[trim=0.8in 0.5in 0.5in 0in, width=3in]{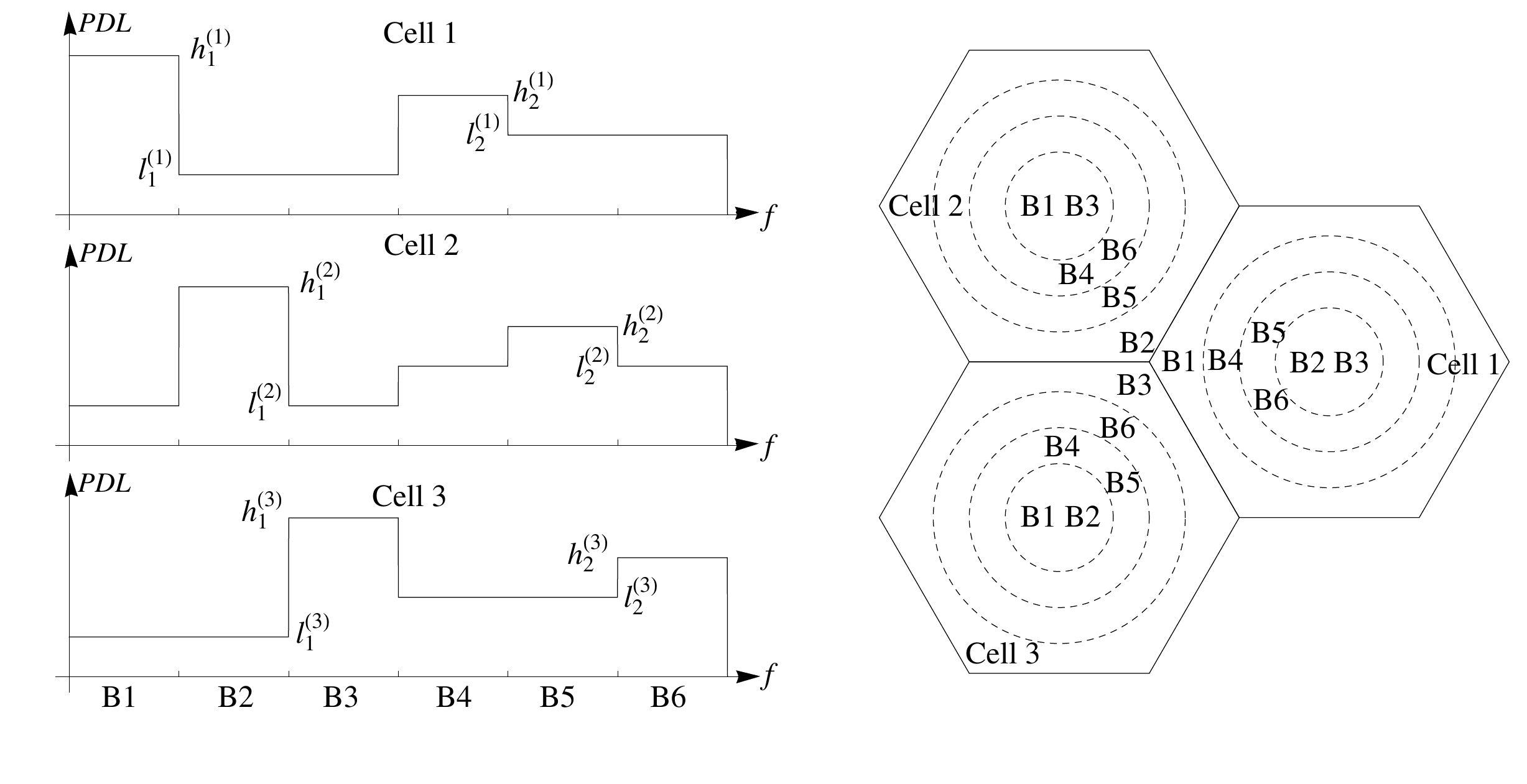}\\
(a)     \qquad   \qquad  \qquad   \qquad\qquad  (b)
\caption{Power density upper limit and coverage in a SFR-4 scheme.}
\label{Fig-ML-SFR}
\end{center}
\end{figure}

SFR has been widely explored and deployed in 4G commercial networks \cite{LTEisNOW}.  Lately, multi-level soft frequency reuse (MLSFR) scheme was presented in \cite{MLSFRYang}. A four-level SFR scheme is illustrated in Fig. \ref{Fig-ML-SFR}, which is the combination of two two-level SFR schemes.  It is worth noticing the specific PDL setting in which the highest PDL pairs with the lowest, the second highest pairs with the second lowest, and so on. In such a way, the interference pattern is further optimized to achieve even better cell edge and overall spectrum efficiency  \cite{MLSFRYang}. 

\section{Discussions}

To cope with the inter-cell interference problem, SFR became an indispensable component of modern cellular system. Under such a circumstance,  the primary band is generally allocated to  cell edge user with low channel gain, and secondary band is assigned to cell centre users with high channel gain. If users are uniformly  distributed,  it is worth noticing that, there are more cell edge users than cell edge ones. Moreover,  the bandwidth is bigger and channel gain is better for cell centre users,  so the data rate provided to the centre users is usually much bigger than edge users. 

In NOMA under SFR, the signals of centre and edge users can be overlapped on primary band. However, such an action is detrimental to fairness by increasing the cell centre data rate and decreasing the cell edge data rate. The capacity gain brought by NOMA to cell centre users is of little meaning since data rate is already more than enough, while the cost of cell edge data rate, even only by 10\%, will significantly harm the user experience.  

Obviously, downlink NOMA will increase the cost of mobile station greatly, since the mobile station should decode not only its own signal, but also signals of other users.  At the same time, this will harm the network security seriously. 

\section{Conclusion}

Although NOMA has larger capacity region than OMA in  downlink, it is detrimental to user fairness in the circumstance of SFR. It also increases the cost of mobile station and harm the network security. So NOMA is not an option for 5G downlink.

\bibliographystyle{IEEEtran}
\bibliography{IEEEfull,MAC}

\end{document}